\documentclass[aps,prc,twocolumn,superscriptaddress,showpacs,amsmath,amssymb,nofootinbib,floatfix]{revtex4-1}
\usepackage{graphicx}  
\graphicspath{ {figures/} } 
\usepackage{bm}     
\usepackage{amssymb}   
\usepackage{amsmath}   
\usepackage{hyperref}
\usepackage{ulem}
\usepackage{natbib}
\usepackage{footmisc}
\usepackage{xcolor}
\usepackage{comment}
\usepackage{lipsum}
\begin{document}
\title{Collective Enhancement of Nuclear Level Density and its fade-out in $^{161}$Dy}
\newcommand{\npd}{Nuclear Physics Division, Bhabha Atomic Research Centre, Mumbai - 400085, INDIA}
\newcommand{\hbni}{Homi Bhabha National Institute, Anushaktinagar, Mumbai- 400094, INDIA} 
\author{T.~Santhosh}\email{tsanthu@barc.gov.in}\affiliation{\npd}\affiliation{\hbni}
\author{P.~C.~Rout}\email{prout@barc.gov.in}\affiliation{\npd}\affiliation{\hbni}
\author{S.~Santra}\affiliation{\npd}\affiliation{\hbni}
\author{A.~Pal}\affiliation{\npd}
\author{G.~Mohanto}\affiliation{\npd}
\author{Ramandeep~Gandhi}\affiliation{\npd}\affiliation{\hbni}
\author{A.~Baishya}\affiliation{\npd}\affiliation{\hbni}
\date{\today}

\begin{abstract}
The nuclear level density is a fundamental quantity in nuclear physics, governing various nuclear reactions and astrophysical processes. In this study, we report on the collective enhancement of nuclear level density and its fade-out with excitation energy in the deformed $^{161}$Dy, obtained through an exclusive measurement of neutron evaporation spectra. The $^{162}$Dy nucleus was populated via the transfer of a triton in a $^{7}$Li-induced reaction on $^{159}$Tb. Statistical model analysis of the neutron spectra revealed a large collective enhancement factor of 42$\pm$2, consistent with microscopic calculations. This enhancement factor is similar to the one obtained for mass A$\sim$170 in our previous measurement. The energy-dependent collective enhancement over a wide range of excitation energies was inferred by combining the present results with the available Oslo level density below the neutron binding energy.
\end{abstract}

\maketitle
The atomic nucleus poses a complex many-body problem, compounded by a limited understanding of the fundamental inter-nucleon interaction. To establish a correlation between experimental data and develop a more comprehensive theory, simplified models such as the shell model and the rotational model have been proposed~\cite{Mayer,BM01}. The equilibrium shape of a nucleus is deformed due to its shell structure, which, in turn, linked to its rotational degrees of freedom. Given the complexity of nuclear forces and the large number of degrees of freedom involved, symmetry properties play a crucial role in characterizing nuclear states. One important aspect of nuclear states is the nuclear level density (NLD), defined as the number of levels per unit MeV energy, is crucial for predicting and interpreting the outcomes of many nuclear reactions. The NLD plays a key role in describing the thermodynamic properties of an excited nucleus, such as temperature and entropy, as well as the decay probabilities of particle emissions from compound nuclear processes using statistical models~\cite{TE}. This makes NLD relevant in estimating low energy astrophysical reaction rates\cite{TR1,TR}, studying giant resonances\cite{DRC1}, and determining reaction rates relevant to energy and isotope production in medical applications\cite{AN}.

Understanding NLD involves investigating various aspects. These include the role of shell effects and nuclear deformation\cite{SKK,BBM}, temperature-dependent behavior, the influence of nuclear structure and pairing correlations\cite{NV}, the impact of nuclear reactions, and the connection to astrophysical processes. By studying these factors, we can gain insights into the fundamental nature of NLD and its implications in nuclear physics and astrophysics.

By considering the nucleus as a collection of non-interacting fermions, Bethe~\cite{HB} derived the level density formula and expressed it as,
\begin{equation}
\rho(E_x) = \frac{1}{12\sqrt{2}{\sigma}}  \frac{\exp{\left({2
\sqrt{aE_x}}\right)}}{a^{1/4}E_{x}^{5/4}},
\end{equation}
where $E_{x}$ is the excitation energy of the nucleus and \textbf{$a$} is the level density parameter with  $a = \pi^2g/6$, \textbf{$g$} is the single-particle level density evaluated at the Fermi energy.
However, this expression does not consider the observed variations in level densities for different types of nuclei (even-even, odd-even, even-odd, and odd-odd).
Bethe formula is a good approximation for explaining the NLD at high excitation energy. However, it becomes inadequate at low energy due to the dominance of nuclear structure effects, such as shell effects and collective excitations, and pairing correlations. Several subsequent works have aimed to include these effects in the level density parameter \textbf{$a$} and excitation energy \textbf{$E_{x}$}. The correction in excitation energy is referred to as the back-shifted Fermi gas (BSFG) model ~\cite{WD}. Another widely used model for describing experimental NLD at low energy is the constant temperature model\cite{GC}. More recently, microscopic models such as HF-BCS ~\cite{PD}, HF-Bologolyubov with combinatorial method ~\cite{SH}, and shell model Monte Carlo~\cite{AO,YA}, have been successful in accurately predicting NLD at low energy ranges, taking into account correlations and structure effects.

To gain more insight into nuclear rotational motion, one can investigate NLDs at excitation energies where statistical principles can be utilized. If collective rotational motion is present, it suggests that there are more degrees of freedom for low-energy excitations, which could result in a substantial increase in the total NLD.

Collective effects are a natural occurrence when extracting microscopic level density. However, phenomenologically, these effects can also be introduced explicitly by incorporating collective enhancement factors into an intrinsic Fermi gas level density. For a deformed nucleus, a large contribution from collective states, known as collective enhancement, is anticipated in addition to the intrinsic level density. The total level density for a deformed nucleus can be expressed as
\begin{equation}
 \rho_{tot} = \rho_{int} K_{vib} K_{rot}
\end{equation}
where $K_{vib}$ and $K_{rot}$ represent the vibrational and rotational enhancement factors, respectively. The contribution from vibrational states is limited due to their large energy spacings, accounting for only a small factor of $\sim 2$ of the total level density in a fully deformed nucleus.

To obtain the level density for a specified spin J of an axially symmetric deformed nucleus, the intrinsic states with specified K (spin projection on the symmetry axis) are summed over,
\begin{equation}
\rho(E_{X},J)=\sum_{K=-J}^{J}\frac{1}{\sqrt{8\pi}\sigma_\perp}e^{{\frac{-K^2}{2\sigma^{2}}}}\rho_{int}(E_{X}-E_{rot})
\end{equation}
This level density was obtained in the earlier works of Bohr, Mottelson, and Bjørnholm for an axially symmetric deformed nucleus and resulted in a factor of $\sigma^{2}_\perp$ higher than for a spherical nucleus. $\sigma_\perp$ is the perpendicular spin cut-off parameter that describes the width of the spin distribution along the perpendicular axis. For Lanthanides, $\sigma_\perp \approx 11 \sqrt{T(MeV)}$, which implies a rotational enhancement factor of around 100 at nucleon binding energies. 
 
As the excitation energy of the nucleus increases, the strength of the coupling between the intrinsic and collective states also increases. However, this coupling eventually dilutes the collective nature of the levels over the neighboring intrinsic states. This leads to the fading out of the collective enhancement at high excitation energies. Experimental studies have been carried out to observe this phenomenon, including measurements of nuclear level densities, gamma-ray strength functions, and other related observables. 

In the past, Jhungans{\it { et al.}}~\cite{Ajung} in a projectile fragmentation experiment, pointed the necessity of inclusion of collective enhancement to explain the observed experimental data. On the contrary, in the work of Komarov {\it { et al.}}~\cite{Skom}, no fade-out of enhancement was observed in the region studied. Recently many experimental observations \cite{KB,GM,DP} confirmed the collective enhancement of NLD and its fade-out although the measured magnitude of enhancement doesn't match with the theoretical predictions. The observed collective enhancement factor is roughly about $\sim$8-10 and fade-out happened around $\sim$14 MeV. 

In our recent measurement \cite{TS}, we have inferred the collective enhancement of NLD in the deformed $^{171}$Yb nucleus, along with its fade-out with excitation energy. This inference was made through an exclusive measurement of neutron spectra. The statistical model analysis of these spectra revealed a significant collective enhancement factor of 40$\pm$3, which is consistent with recent microscopic model predictions but stands out as an anomalous result compared to measurements in nearby deformed nuclei. 
Importantly, we highlight the significance of the large collective enhancement in the radiative neutron capture cross section, which holds astrophysical relevance. These findings emphasize the importance of further experimental investigations to refine our understanding of collective enhancement in NLD and its potential impact on nuclear reactions.

\begin{figure}[h]
    \centering
    \includegraphics[width=0.4\textwidth]{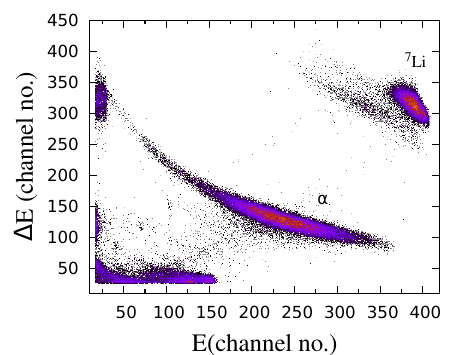}
    \caption{A standard 2D plot illustrating the energy loss of particles in one of the strips of a $\Delta E$-$E$ telescope.}
        \label{fig:01}
\end{figure}

\begin{figure}[t]
    \centering
    \includegraphics[width=0.5\textwidth]{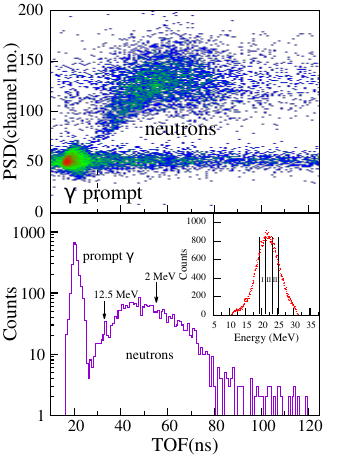}
    \caption{(top)~The pulse shape discrimination (PSD) as a function of time of flight (TOF) in the $^7$Li + $^{159}$Tb reaction. The prompt $\gamma$-rays and neutrons are clearly separated.\\ (bottom) A typical time of flight(TOF) in the $^7$Li + $^{159}$Tb reaction, specifically for the central energy bin of alpha particles. Two representative neutron energies are indicated by arrows, highlighting their positions in the TOF spectrum.  The inset in the figure displays the projected alpha energy spectrum, with three energy bins indicated for reference.}    
    \label{fig:02}
\end{figure}
The present work focuses on the determination of the nuclear level density of $^{161}$Dy, which is produced through an incomplete fusion reaction using weakly bound $^{7}$Li projectiles. The $^{7}$Li nucleus has a cluster structure, with a dominant reaction mechanism of breakup or transfer capture of one of the clustered parts to the target nucleus. In this case, the $^{7}$Li nucleus is mainly composed of an $\alpha$ particle and a triton, and triton transfer or breakup-fusion is the dominant process during incomplete fusion reaction. The compound nucleus $^{162}$Dy was populated by capturing a triton in the $^{159}$Tb nucleus. The NLD was then extracted by measuring the evaporated neutron spectra from $^{162}$Dy.

The experiment was carried out at the BARC-TIFR 14UD Pelletron laboratory at Mumbai. A pulsed $^{7}$Li beam (width $\sim$1.5~ns (FWHM) and period $\sim$107~ns) with an energy of 40 MeV was directed at a self-supported $^{159}$Tb target, which had a thickness of 2.8 mg/cm$^2$. The target thickness was determined using weighing and was mounted on stainless steel frames. A blank frame was also employed to estimate any scattering caused by the beam hitting the target ladder. A current integrator was utilized to monitor the total beam incident on the target.

In order to detect the outgoing alpha particles in the break-up/transfer channel of $^{7}$Li, two $\Delta$E-E  telescopes(5~cm$\times$5~cm) were employed, each consisting of a double-sided silicon strip detector (DSSD). The telescopes were positioned at a distance of 10 cm from the target center, at an average angle of $\pm$60$^\circ$ relative to the beam direction. The $\Delta$E and E detectors had thicknesses of 50$\mu$m and 1500$\mu$m, respectively. Each DSSD had 16 strips on each side and covered an angular range of 25$^\circ$. Alpha particles were identified using Bethe-Bloch energy loss technique in strip telescopes, these alpha particles were then used to find the neutrons in coincidence.

Evaporated neutrons from the compound nucleus were detected using an array of 15 liquid scintillation(LS, EJ301) detectors arranged in circular geometry~\cite{pcr2}. The detector array was placed 70 cm from the target center at angles ranging from 58$^\circ$ to 143$^\circ$ with respect to the incident $^{7}$Li beam. The standard pulse-shape discrimination(PSD) technique was used for unambiguous detection of neutrons against gamma rays. Neutron energy spectra were determined using the time-of-flight (TOF) method, which involved measuring the time taken for neutrons to travel from the target to the detectors using the pulsed beam bunched with a period of approximately 107 ns.

All signal readouts from LS detectors and strip detectors had been recorded in list mode using a VME based data acquisition system. A shadow pyramid bar of iron plates was kept in front of the LS detector array to estimate the scattered neutron contribution from the surroundings. The beam dump was shielded with borated paraffin and lead blocks to reduce the background.
 
\begin{figure}[t]
\begin{center}
\includegraphics[width=0.45\textwidth]{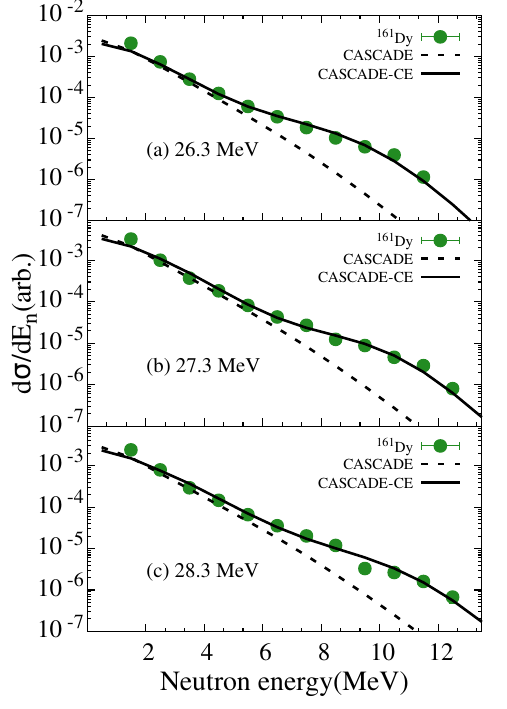}
\end{center}
\caption{Comparison of neutron spectra with Statistical Model calculation using the level density parameter A/8.5 MeV$^{-1}$. Solid line show calculation with collective enhancement (CASCADE-CE) and dashed line is without collective enhancement (CASCADE). Three excitation energies (a) 26.3~MeV, (b) 27.3~MeV and (c) 28.3~MeV, respectively, for three alpha energy gates(III, II and I) as shown in inset of the Figure~\ref{fig:02}.}
\label{fig:ns}
\end{figure}

\begin{figure}[t]
\begin{center}
\includegraphics[width=0.43\textwidth]{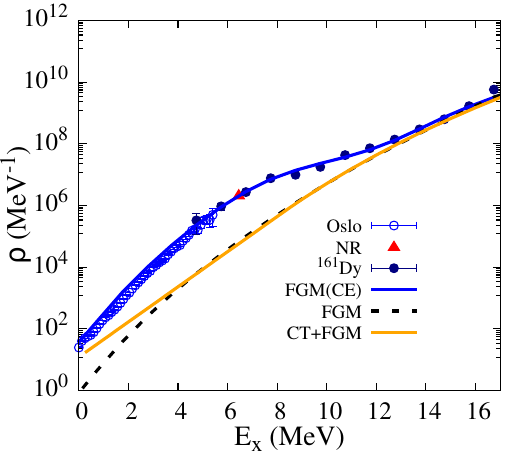}
\end{center}
\caption{The normalized level density as a function of excitation energy is depicted in the plot, with Oslo data represented by open circles~\cite{AS} and the present experiment by filled circles. The NLD is normalized to the level density at the neutron resonance point (NR), indicated by a triangle. The dashed red line represents the intrinsic level density from the Fermi gas model (FGM), while the blue solid line represents the Fermi gas level density with collective enhancement (FGM(CE)). Solid orange line shows the level density from constant temperature + Fermi gas model(CT+FGM).}
\label{fig:rho}
\end{figure}

A typical 2D plot of the energy loss of particles in the $\Delta$E and E strip detectors for one of strips is shown in Figure~\ref{fig:01}. It is observed that the alpha particles are well separated from $^7$Li. The PSD in relation to the TOF for the $^7$Li + $^{159}$Tb reaction is displayed in the top panel of Figure~\ref{fig:02} , while the bottom panel depicts the TOF spectra gated with alpha particles. To extract the absolute neutron TOF, the prompt gamma signal illustrated in Figure ~\ref{fig:02} was used as a reference. The inset in Figure~\ref{fig:02} shows the alpha energy distribution peaking at $\sim$~22~MeV, which corresponds to the energy of the beam velocity. Different energy bins were identified to obtain the neutron spectra for various excitation energies. The neutron TOFs were subsequently converted into neutron energy spectra using appropriate Jacobian factor. The efficiency of neutron  detectors as a function of incident energy and threshold were estimated using a Monte Carlo simulation~\cite{pcr2}, which was validated using measured neutrons from $^7$Li(p,n) reaction. The efficiency corrected neutron energy spectra after converting in to the center of mass are shown in Figure~\ref{fig:ns}. In the center of mass frame, it is also observed that the forward and backward spectra exhibit symmetry around $90^\circ$. This symmetry implies that the emission of neutrons originates from a statistically equilibrated system.

In order to quantify the collective enhancement, if at all is reflected in the neutron spectra, we compared the experimentally measured neutron spectra with statistical model code CASCADE\cite{puhl}. The code has been very successful in explaining the evaporation spectra. The following E$_X$ and J dependent level density expression has been used in the CASCADE code, 
\begin{equation}
\rho(E_x,J)=\frac{(2J+1)\sqrt{a}}{12U^2}\left(\frac{\hbar^2}{2\Im}\right)^{3/2}e^{2\sqrt{aU}},
\end{equation}
where $U$ is defined as $U = E_X - E_{\rm{rot}} - \Delta_P$, with $\Delta_P$ being the pairing energy calculated using $\Delta_P=\frac{12}{\sqrt{A}}$ and $E_{\rm{rot}}$ representing the rotational energy given by $E_{\rm{rot}} = \left(\frac{\hbar^2}{2\Im}\right) J(J+1)$, where $\Im$ denotes the moment of inertia defined as $~ \Im = I_0\left(1 + \delta_1 J^2 + \delta_2 J^4\right)$ with $I_0$ being the rigid body moment of inertia, and $\delta_1$, $\delta_2$ representing the deformability parameters of a liquid drop nucleus\cite{SC}. 

The expression for the level density parameter $a$ has been parameterized using the Ignatyuk prescription~\cite{avi} as,
$a = \tilde{a} \left[1-\frac{\Delta_S}{U}(1 - e^{-\gamma U})\right]$
where $\tilde{a}$ represents the asymptotic value of the NLD parameter in the liquid drop region. $\Delta_S$ denotes the shell correction energy, which is calculated as the difference between the experimental binding energy and the binding energy calculated from the liquid drop model (LDM). $\gamma$ represents the damping parameter\cite{WD}.

The CASCADE code has the feasibilty to include collective enhancement($K_{coll}$) explicitly, and is modeled as below, 
\begin{equation}
 K_{coll} = 1+A_{en} ({1+\exp[(E-E_{cr})/d_{cr}]})^{-1}
 \end{equation}
where A$_{en}$ represents the maximum factor by which enhancement is collectively increased, E$_{cr}$ denotes the energy at which the enhancement decreases to half of its maximum value, and d$_{cr}$ represents the width of the transition region.

The excitation energy of $^{162}$Dy populated in the present case was such that the high energy neutrons($>$ 5MeV) is predominantly from first step neutron emission. {\textcolor{blue}{ Above 6 MeV, the first step contribution is more than 80\%, and above 8 MeV, it is 100\%}}. This implies that the collective enhancement extracted in this measurement is for $^{161}$Dy.

Although the statistical model do not consider many details of the nuclear interactions, the empirical parameters determined by the fitting of experimental data reflect the collective enhancement, shell effects, the pairing effects, etc. The Figure ~\ref{fig:ns} presents a comparison of neutron spectra obtained at three different excitation energies with CASCADE results. The dashed lines(CASCADE) in the figure represent CASCADE calculations without considering collective enhancement, while solid lines includes this effect(CASCADE-CE). The CASCADE statistical model calculation matches the low-energy neutron part, which has a constant slope, and collective enhancement is added to reproduce the experimental results. This is achieved by adjusting the inverse level density parameter $k$(defined as $k=A/a$) and collective enhancement function parameters ($A_{en}$, $E_{cr}$, and $d_{cr}$) in the CASCADE input. The inverse level density parameter k = A/8.5 MeV$^{-1}$ was used in the calculation. The value k was determined by fitting to the low-energy neutron spectra ($<$ 6 MeV), and then the enhancement function parameters were subsequently modified to reproduce the experimental data. The best parameters were determined by simultaneously fitting the three excitation energies, yielding a maximum collective enhancement factor of 42$\pm$2, and values of $E_{cr}$ and $d_{cr}$ at 8.5$\pm$0.5 MeV and 1.2$\pm$0.2 MeV, respectively.

Our primary objective is to determine the excitation energy-dependent level density, which cannot be fully extracted from neutron spectra without being influenced by a particular model. To address this issue, a model-independent method has been developed to obtain the total level density. The initial step involves fitting the center-of-mass neutron spectra to a CASCADE calculation that employs the prescribed level density(in the present case equation [4]). Then using this NLD prescription with collective enhancement inclusion we looked for the optimal value of the level density parameter $a$ that provides the best fit to the experimental data. Once these optimal fits are obtained, the total NLD for the residual nucleus following the emission of one neutron at a specific excitation energy is extracted using the below expression,
\begin{equation}
    \rho(E_x)\propto \frac{(d\sigma/dE_{n})_{M}}{(d\sigma/dE_{n})_{T}}\rho_{T}(E_x)
\end{equation}
The $\rho_{T}(E_x)$ was determined by adding up the NLDs that best fit the data for all values of angular momentum. ${(d\sigma/dE_{n})_{M}}$ and ${(d\sigma/dE_{n})_{T}}$ are the measured and cascade-predicted differential neutron cross sections, respectively. 
The above scaling technique was described in Ref~\cite{DRC}, and it is based on the "first step hypothesis". This hypothesis posits that the high-energy particles observed in the spectrum at intermediate excitation energies of the compound nucleus primarily result from the first-step decay. 

\begin{figure}[t]
\begin{center}
\includegraphics[width=0.5\textwidth]{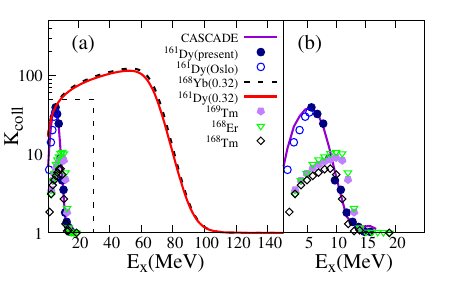}
\end{center}
\caption{(a) The figure illustrates the enhancement factor as a function of excitation energy, obtained by comparing the level density calculations incorporating Kcoll (collective enhancement) in the CASCADE model to the constant temperature + Fermi gas model (CT+FGM), referred to as CASCADE. It also presents the combined enhancement extracted from Oslo data along with the present measured enhancement. The figure further includes experimental enhancement factors for various nuclei~\cite{DP}. Additionally, the enhancement function extracted from Hansen and Jensen's prescription~\cite{HJ} is showcased for the nuclei $^{168}$Yb and $^{161}$Dy which have $\beta_{2}$ = 0.32. (b) Represents the zoomed part of the Fig.~(a) marked in dashed box.}
\label{fig:kcoll}
\end{figure}

In order to derive the absolute total nuclear level density of $^{161}$Dy as a function of excitation energy, the relative nuclear level density $\rho(E_x)$ was modified by normalizing it with the level density 2.1$\times 10^{6}~MeV^{-1}$ obtained from neutron resonance data\cite{NR} at the neutron binding energy 6.45 MeV. The resulting data was then merged with the level density data obtained via the Oslo method to generate a comprehensive overview covering the energy range from above zero to 16 MeV. Figure~\ref{fig:rho} illustrates that the level density increases as the excitation energy increases, with complete fade-out occurring around 15 MeV.
 
The Figure~\ref{fig:kcoll} shows the enhancement factor as a function of excitation energy extracted from the ratio of level density with Kcoll (collective enhancement) included in the CASCADE calculations to the Fermi gas model + constant temperature model(CT+FGM), denoted as CASCADE. The enhancement extracted from Oslo data combined with present measured enhancement is also shown. These are extracted by taking the ratio between experimental level density and CT+FGM level density. Experimental enhancement factors extracted for various nuclei are plotted as well. Additionally, the enhancement function extracted from the Hansen and Jensen's prescription~\cite{HJ} for the nuclei $^{168}$Yb and $^{161}$Dy is included. The plot provides insights into the collective enhancement behavior with respect to excitation energy, highlighting the significance of different modeling approaches and experimental measurements in understanding the underlying physics of nuclear excitations.

The potential sources of uncertainty in the current measurement are outlined as follows. In cases where the target material contains impurities such as carbon and oxygen, there may be a significant presence of background neutrons. Although these impurities may only constitute a small fraction of the material's weight, their impact can be amplified by the characteristics of the neutron spectra for light targets and kinematic focusing at forward angles. As a result, measurements are often limited to backward angles, especially for high-energy neutron spectra of interest. However, this restriction does not affect the determination of the NLD as pre-equilibrium nuclear reactions are more significant at forward angles and are not as relevant at low beam energies, as is the case in the current measurement. In the present measurement, although the triton breakup/transfer-fusion reaction is expected to be the primary contributor to the coincident neutron spectra, there are other direct processes that could also play a role. However, we can safely disregard the proton pickup and two-neutron transfer cross sections as their contribution is negligible\cite{SKP}. The most significant reaction to consider is the deuteron transfer followed by $^5{He}$ breakup. However, since the spectroscopic factor for the d + $^5{He}$ configuration is anticipated to be much smaller than that of the t + $^4{He}$ configuration, the contribution of this reaction is insignificant\cite{pcr1}.

In conclusion, our study utilized an incomplete fusion reaction to investigate the compound nucleus of $^{162}$Dy and observed the resulting neutron evaporation spectra. By employing the  statistical model calculations and comparing with experimental data, we found a significant deviation between standard statistical model calculations and experimental results, indicating the presence of collective enhancement. To address this discrepancy, we proposed an excitation energy-dependent collective enhancement based on Hansen and Jensen's suggestion. Incorporating this collective enhancement, we were able to successfully explain the experimental data. Furthermore, our statistical model analysis revealed a maximum collective enhancement of 42$\pm$2, with fade-out observed at around 15 MeV. These results are in agreement with the recent microscopic calculations. Recent theoretical works on NLD using finite-temperature relativistic Hartree Bogoliubov model \cite{jz}, showed that the enhancement in the mass region $A$ = 160-170 is $\approx$ 40. In another work~\cite{smg}, using a microscopic level density model, a similar magnitude of collective enhancement has been predicted.  The energy-dependent collective enhancement was inferred by combining our results with Oslo level density data below the neutron binding energy, providing insights into the behavior of collective enhancement over a wide range of excitation energies. These findings highlight the importance of considering collective enhancement in statistical model calculations for describing compound nucleus reactions and provide valuable information for further studies in nuclear reaction dynamics. 
 
We thank B. K. Nayak and D.R. Chakraborty for their valuable comments and suggestions. We thank the PLF staff for smooth operation of the accelerator. The author, Santhosh(IF180174) is sincerely grateful to DST for financial support under the DST-INSPIRE Fellowship scheme.

\end{document}